\documentclass[12pt]{article}
\usepackage{amsmath,latexsym}
\usepackage{graphicx}
\begin{document}

 \title{\Huge Discussion on some characteristics of the  Charged Brane-world Black holes}
 \author{ M.Kalam$^{\dag}$, F.Rahaman$^*$, A.Ghosh$^*$ and B. Raychaudhuri$^{\ddag}$  }
\date{}
 \maketitle

 \begin{abstract}
Several physical natures of charged brane-world black
 holes have been investigated. At first,
 time-like and null geodesics of the charged brane-world black
 holes are presented. We also analyze all the possible motions by
 plotting the effective potentials for various parameters for circular and radial geodesics.
Secondly,  we  investigate the motion of test particles in the
gravitational field of charged brane-world black holes  using
Hamilton-Jacobi ( H-J ) formalism.
 We have considered charged and
 uncharged test particles and examine its behavior both in static
 and non-static cases. Thirdly, thermodynamics of the charged brane-world
 black holes are studied. Finally, it has been also shown that there is no
 phenomenon of superradiance for an incident massless scalar field
 for this black hole.
\end{abstract}
  \footnotetext{ Pacs Nos:  04.20 Gz, 04.50 + h, 04.20 Jb   
\\
$\dag$Dept.of Phys.,Netaji Nagar College for Women, Regent
Estate, Kolkata-700092, India: \\
E-Mail:mehedikalam@yahoo.co.in \\
 $*$Dept.of Mathematics, Jadavpur University, Kolkata-700 032, India: \\
 E-Mail:farook\_rahaman@yahoo.com\\
 $\ddag$Dept. of Phys. , Surya Sen Mahavidyalaya, Siliguri, West Bengal, India :\\
 E-mail: biplab.raychaudhuri@gmail.com
 }
    \mbox{} \hspace{.2in}
\begin{center}
\title{\Large 1.  \textsc{Introduction} }
\end{center}
 In recent, scientists have given their attention to brane world gravity. In brane world models, the ordinary matter fields are confined on a three
 dimensional subspace, called brane embedded in 1+3+d dimensions in which the gravity can propagate in the d-extra dimensions. Here, the
 d-extra dimensions need not all be small or even compact. Most of the recent studies consider a simple version of the brane world scenario where
 all matters ( except gravity ) are confined to a 3-brane embedded in a five dimensional space-time ( bulk ) while gravity can propagate in the bulk.
 Recently, Dadhich et. al [1] have presented a spherically symmetric solution which describes a black hole localized on a three brane in five dimensional gravity
 in the brane world scenario. This black hole ( without electric charge ) is termed as tidal charged black hole. In this case, tidal charge is arising
 via gravitational effects from the fifth dimension i.e. it is arising from the projection on to the brane of free gravitational field effects
 in the bulk. Chamblin et. al [2] studied charged brane world black holes in Randall and Sundrum model. In this model, they assumed our universe
 as a domain wall in asymtotically anti-de Sitter space. This type of black holes can have two types of "charge", one comes from the bulk Weyl tensor and
  the other from a gauge field trapped on the wall. By using the brane-world Einstein equations, a $Reissner-Norstr\ddot{o}m$ (RN) geometry can be
  found on the domain wall provided that only the bulk Weyl charge is present [3]. Chamblin et. al showed that the extent of the horizon in the
  fifth dimension for a charged black hole is usually less than for an uncharge black hole that has the same mass or the same horizon radius on the wall.   \\

In this paper, we will discuss the behavior of the time-like and
null geodesics of the charged brane-world black holes.We will
analyze all the possible motions by plotting the effective
potentials for various parameters for circular and radial
geodesics. Also we will investigate the motion of test particles
in the gravitational field of charged brane-world black holes
using Hamilton-Jacobi method. We have considered charged and
 uncharged test particles and examine its behaviour both in static
 and non-static cases. Thermodynamics of the charged brane-world
 black holes are studied. It has been also checked  that if there is
 any phenomenon of superradiance or not for an incident massless scalar field
 for this black hole.
 \\

\begin{center}
\title{\Large2. \textsc{Charged Brane-World Black holes metric} }
\end{center}

 A Charged Brane-World Black holes  metric can be written as[2]
\begin {equation}
        ds^2=  - f(r) dt^2 +  \frac{dr^2}{f(r)} + r^2 (d\theta^2 + sin^2\theta d\phi^2 )
  \end{equation}
where
\begin{eqnarray*}
         f(r) = 1- \frac{2GM}{r}+\frac{Q^2 + \beta}{r^2} +
\frac{l^2 Q^4}{20 r^6}
 \end{eqnarray*}
M,Q and $\beta$ corresponds to Mass,electro-magnetic charge and
tidal charge of the black hole respectively.\\
The electric gauge potential have the form $ A_i = - \Phi(r) dt $
with $\Phi(r)=\frac{Q}{r}$.
\begin{center}

\pagebreak

\title{\Large 3.  \textsc{The Geodesics} }
\end{center}

Let us now write down the equation for the geodesics in the
metric (1) . From
\begin{equation}
               \frac{d^2 x^\mu}{d\tau^2} + \Gamma^\mu_{\nu\lambda}
               \frac{d x^\nu}{d\tau}\frac{d x^\lambda}{d\tau}=0
               \end{equation}
we have
\begin{equation}
               \frac{1}{f(r)}\left(\frac{d r}{d\tau}\right)^2 = \frac{E^2}{f(r)} - \frac{J^2}{r^2} -  L
               \end{equation}
\begin{equation}
               r^2\left(\frac{d \phi}{d\tau}\right) =  J
               \end{equation}
\begin{equation}
                \frac{d t}{d\tau} = \frac{E}{f(r)}
               \end{equation}

where the motion is considered in the $ \theta  = \frac{\pi}{2}$
plane and constants E and J are identified as the energy per unit
mass and angular momentum, respectively , about an axis
perpendicular to the invariant plane $ \theta  = \frac{\pi}{2}$.
Here $\tau$ is the affine parameter and L is the Lagrangian having
values 0 and 1, respectively, for massless and massive
particles. \\
The  equation for radial geodesic ( $ J =0$):

\begin{equation}
        \dot{r}^2 \equiv \left(\frac{dr}{d\tau}\right)^2 = E^2 - L f(r)
\end{equation}

Using  eqn.(5) and  eqn.(3)   we get
\begin{equation}
         \left(\frac{dr}{dt}\right)^2 = \left (1- \frac{2G M}{r} + \frac{Q^2 + \beta}{r^2}+\frac{l^2 Q^4}{20 r^6}\right)^2
         \left[ 1 - \frac{L}{E^2} \left (1- \frac{2G M}{r} + \frac{Q^2 + \beta}{r^2}+\frac{l^2 Q^4}{20
         r^6}\right)\right]
\end{equation}
\begin{center}
\title{\bf 3.1  \textsc{Motion of Massless Particle ( L=0 )} }
\end{center}

In this case,
\begin{equation}
         \left(\frac{dr}{dt}\right)^2 =  \left (1- \frac{2G M}{r} + \frac{Q^2 + \beta}{r^2}+\frac{l^2 Q^4}{20 r^6}\right)^2
\end{equation}
Neglecting the higher order of $ ( Q^2 + \beta )  $ , $ l^2 Q^4$ ,
$ GM $ and after integrating, we get the $t - r $ relationship as

$
          \pm t =  r + 2GM ln r + \frac{Q^2 + \beta}{r}+ \frac{l^2 Q^4}{100
          r^5}- \frac{4 G^2 M^2}{r}- \frac{(Q^2 + \beta)^2}{3r^3}-\frac{l^4 Q^8}{4400
          r^{11}}
       + \frac{GM(Q^2 + \beta)}{r^2}
       - \frac{l^2 Q^4(Q^2 +
        \linebreak
          \beta)}{70r^7}+ \frac{G M l^2 Q^4}{30r^6}$ $.
$ \\

The $t - r$ relationship is depicted in Fig. 1.\\

\begin{figure}[htbp]
   \centering
        \includegraphics[scale=.5]{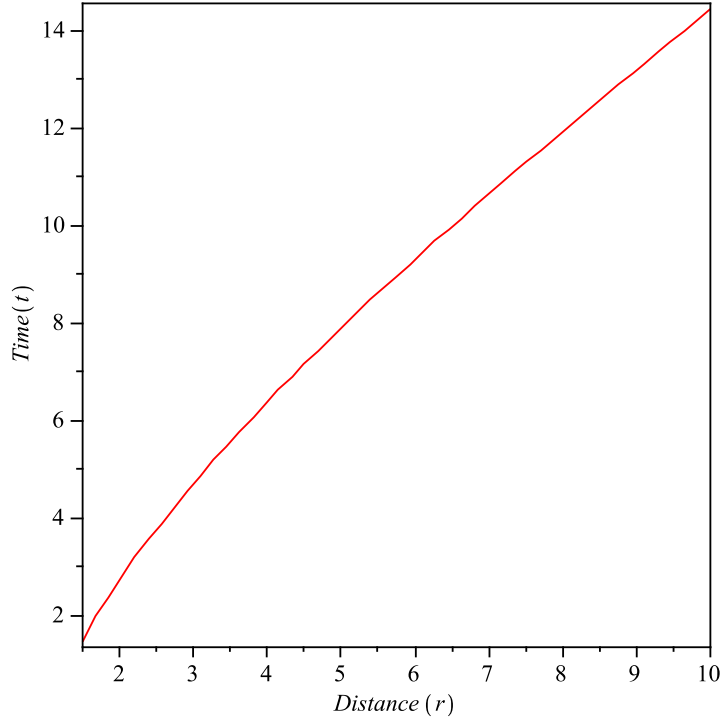}
    \caption{$t - r$ relationship for massless particle( choosing $G= M= Q= l=1$, $\beta =1$ ) }
    \label{fig:Charged Brane black hole}
\end{figure}

Again, from equation (6) we get
\begin{equation}
        \dot{r}^2 \equiv \left(\frac{dr}{d\tau}\right)^2 =  E^2
\end{equation}

After integrating, we get the $\tau - r $ relationship as
\begin{equation}
          \pm E\tau =     r
\end{equation}

We show graphically (see Fig. 2 ) the variation of proper-time ($\tau$) with respect to radial co-ordinates (r) .\\

\begin{figure}[htbp]
    \centering
        \includegraphics[scale=.5]{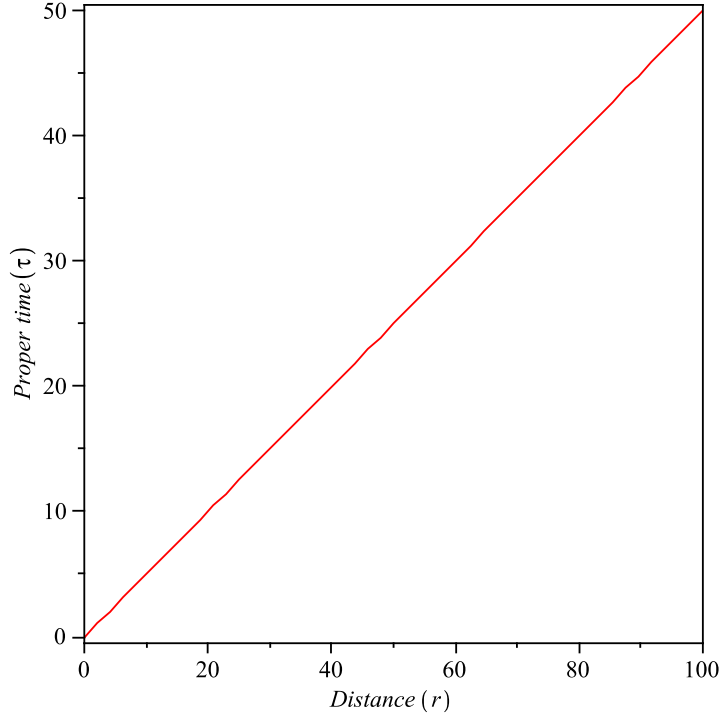}
    \caption{$\tau - r$ relationship for massless particle ( choosing $E =2$ ) }
   \label{fig:Charged Brane black hole}
\end{figure}

\pagebreak
\begin{center}
\title{\bf 3.2  \textsc{Motion of Massive Particles ( L=1 )} }
\end{center}

In this case,
\begin{equation}
         \left(\frac{dr}{dt}\right)^2 = \left (1- \frac{2G M}{r} + \frac{Q^2 + \beta}{r^2}+\frac{l^2 Q^4}{20
         r^6}\right)^2 . \frac{1}{E^2}
         \left(E^2 - 1+ \frac{2G M}{r} - \frac{Q^2 + \beta}{r^2}-\frac{l^2 Q^4}{20 r^6}\right)
\end{equation}
After integrating, we get
\begin{equation}
          \pm t = \int \frac{E dr}{\left (1- \frac{2G M}{r} + \frac{Q^2 + \beta}{r^2}+\frac{l^2 Q^4}{20
         r^6}\right)\sqrt{E^2 - 1+ \frac{2G M}{r} - \frac{Q^2 + \beta}{r^2}-\frac{l^2 Q^4}{20 r^6}   }}
\end{equation}
This gives the $t - r $ relationship as (neglecting the higher
order of $(Q^2+\beta)$ and $l^2 Q^4$)  \\
(see graphical Fig. (3))

 $
          \pm t = \frac{E}{\sqrt{E^2-1}}[ r + \left(2GM -
          \frac{GM}{E^2-1}\right)ln r -
          \left(4G^2M^2-(Q^2+\beta)+\frac{Q^2+\beta}{2(E^2-1)}+\frac{3}{2}\frac{G^2M^2}{(E^2-1)^2}-\frac{2G^2M^2}{E^2-1}\right)\frac{1}{r}
\linebreak
           -\frac{1}{r^2}(4G^3M^3 - 2GM(Q^2+\beta)-\frac{3}{4}\frac{GM(Q^2+\beta)}{(E^2-1)^2}- \frac{5}{4}\frac{G^3M^3}{(E^2-1)^3}+GM(\frac{Q^2+\beta}{2(E^2-1)}
          + \frac{3}{2}\frac{G^2M^2}{(E^2-1)^2})-\frac{GM}{E^2-1}(4G^2M^2-(Q^2+\beta)))+..............       ]$
\begin{figure}[htbp]
   \centering
        \includegraphics[scale=.5]{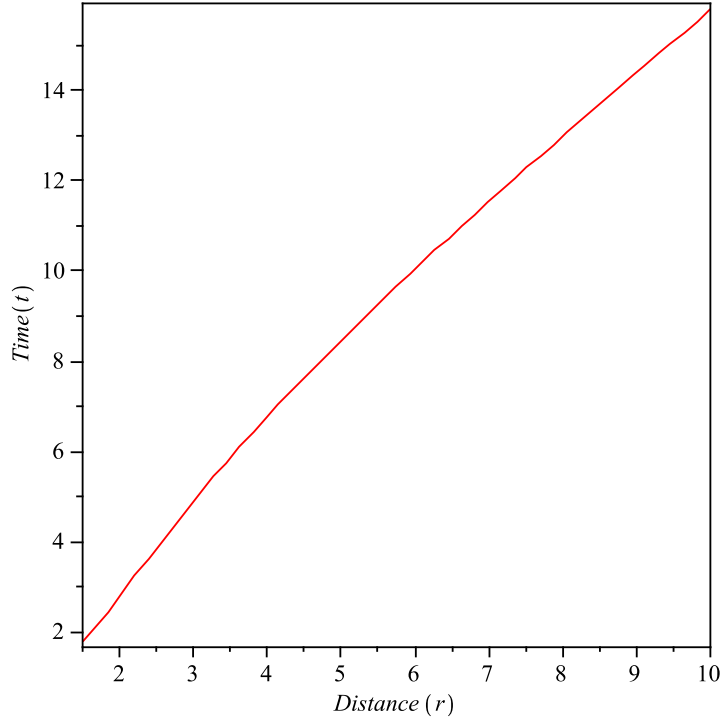}
    \caption{$t - r$ relationship  for massive particle( choosing $G= M= Q= l=1, E=2$, $\beta =1$ )}
    \label{fig:monopole}
\end{figure}

Again, from equation (6) we get
\begin{eqnarray*}
        \dot{r}^2 \equiv \left(\frac{dr}{d\tau}\right)^2 =
        E^2 - \left (1- \frac{2G M}{r} + \frac{Q^2 + \beta}{r^2}+\frac{l^2 Q^4}{20
         r^6}\right)
\end{eqnarray*}

 After simplification, we get

\begin{eqnarray*}
          \pm \int  d\tau = \int \frac{ dr }{\sqrt{E^2- \left (1- \frac{2G M}{r} + \frac{Q^2 + \beta}{r^2}+\frac{l^2 Q^4}{20
         r^6}\right)}}
\end{eqnarray*}

Neglecting the higher order of $ ( Q^2 + \beta )  $ and  $
l^2     Q^4$  gives the $\tau - r $ relationship as

$
         \pm  \tau =  \frac{1}{\sqrt{E^2-1}} [ r-\frac{GM }{E^2-1} ln r-\left(\frac{Q^2+\beta}{2(E^2-1)}+\frac{3 G^2
          M^2}{(E^2-1)^2}\right)\frac{1}{r} + \frac{3}{4} \frac{GM(Q^2 + \beta)}{(E^2-1)^2}\frac{1}{r^2}-\frac{(Q^2+\beta)^2}{8(E^2-1)^2}\frac{1}{r^3}
\linebreak
        - \frac{l^2 Q^4}{200(E^2-1)}\frac{1}{r^5}+\frac{GMl^2Q^4}{80(E^2-1)^2}\frac{1}{r^6}-\frac{3 l^2 Q^4(Q^2+\beta)}{560(E^2-1)^2}\frac{1}{r^7}
        - \frac{3 l^4 Q^8}{35200(E^2-1)^2}\frac{1}{r^{11}}
        ]
$

We show graphically (see Fig. 4 ) the variation of proper-time ($\tau$) with respect to radial co-ordinates (r) .\\

\begin{figure}[htbp]
    \centering
       \includegraphics[scale=.45]{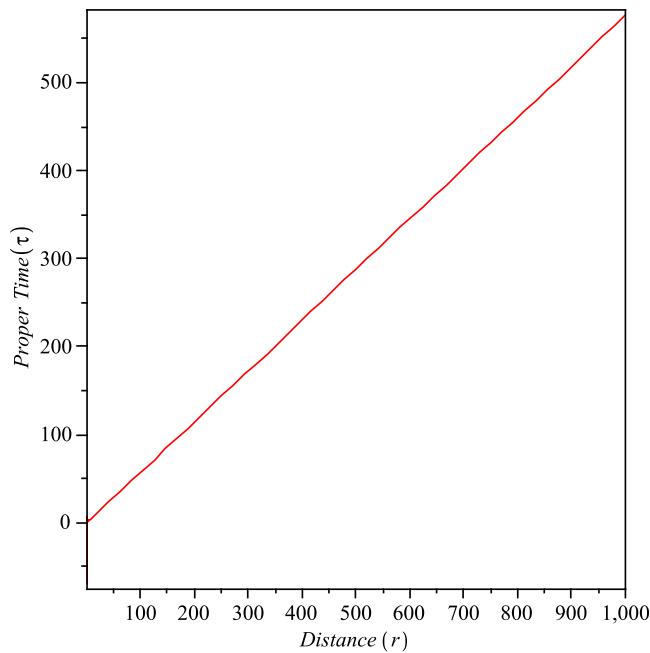}
    \caption{$\tau - r$ relationship for massive particle ( choosing $G= M= Q= l=1$, $\beta =1$ and $E =2$ ) }
    \label{fig:Charged Brane black hole}
\end{figure}

\pagebreak
\begin{center}
\title{\Large 4.  \textsc{EFFECTIVE POTENTIAL} }
\end{center}
From the Geodesic equation (3),(4) and (5) we can write
\begin{equation}
\frac{1}{2}\left(\frac{dr}{d\tau}\right)^2 =
\frac{1}{2}\left[E^2-f(r)\left(\frac{J^2}{r^2}+L\right)\right]
\end{equation}
Comparing eqn.(13) with $\frac{\dot{r}^2}{2} + V_{eff} = 0 $, one
can get the effective potential,which depends on E and L as
follows :
\begin{equation}
V_{eff} =
-\frac{1}{2}\left[E^2-f(r)\left(\frac{J^2}{r^2}+L\right)\right]
\end{equation}
\begin{center}
\title{\bf 4.1  \textsc{For  Massless Particle ( L=0 )} }
\end{center}
At first consider, at the radial geodesics where J=0. The
corresponding $V_{eff}$ is given by
\[
 V_{eff} = - \frac{E^2}{2}
\]
If, $E = 0$, then $V_{eff} = 0$ i.e. the particle behaves like a
"free particle". The graph of $V_{eff}$ for $E \neq 0$ is shown
in Fig. 5. It is obvious that the behaviour of these geodesics is
independent on the charge and mass of the black hole.
\begin{figure}[htbp]
    \centering
       \includegraphics[scale=.5]{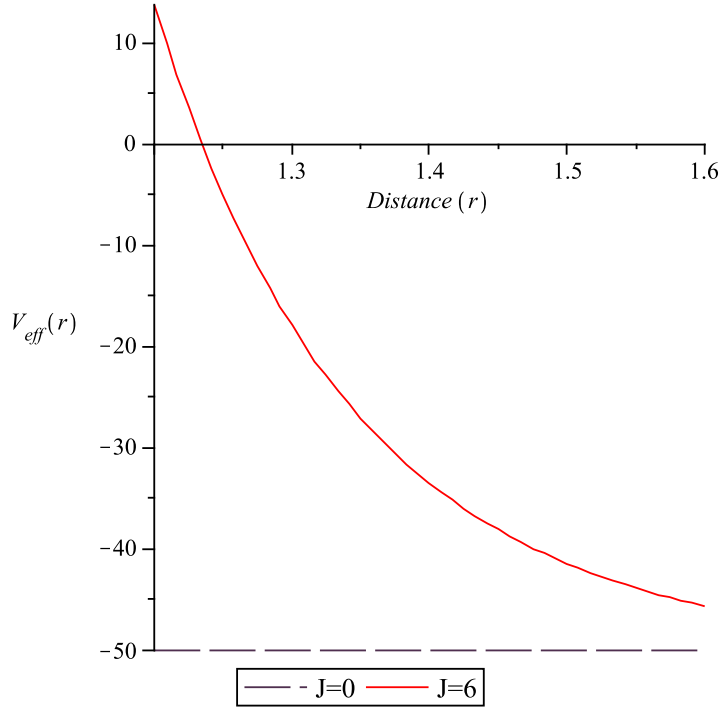}
    \caption{The effective potential for photons for M=2,G=1,$\beta=1$,$Q=\sqrt{2},l^2=80$,E=10. }
    \label{fig:Charged Brane black hole}
\end{figure}
\\

Now consider, for circular geodesics where $ J \neq 0$. The
corresponding effective potential is,
\begin{equation}
 V_{eff} = - \frac{E^2}{2} +
 \frac{J^2}{2r^2}\left(1-\frac{2GM}{r}+\frac{Q^2+\beta}{r^2}+\frac{l^2Q^4}{20r^6}\right)
\end{equation}
For $ r \rightarrow 0 $, the effective potential,$V_{eff} (r)$ is
very large  and approaches $- \frac{E^2}{2}$ when $ r \rightarrow
\infty $. At the horizons, $V_{eff} = - \frac{E^2}{2}$. Let us
consider, the effective potential for  $E =0$ [put E=0 in
eqn.(15)]. The roots of the potential coincide with the horizon
values for this case. The potential is negative between the
horizons. Hence, the particle would be bounded between the
horizons. Again, since potential has a minimum between the
horizons, stable circular orbits do exists. Fig.6 is an example
for such a case.
\begin{figure}[htbp]
    \centering
       \includegraphics[scale=.5]{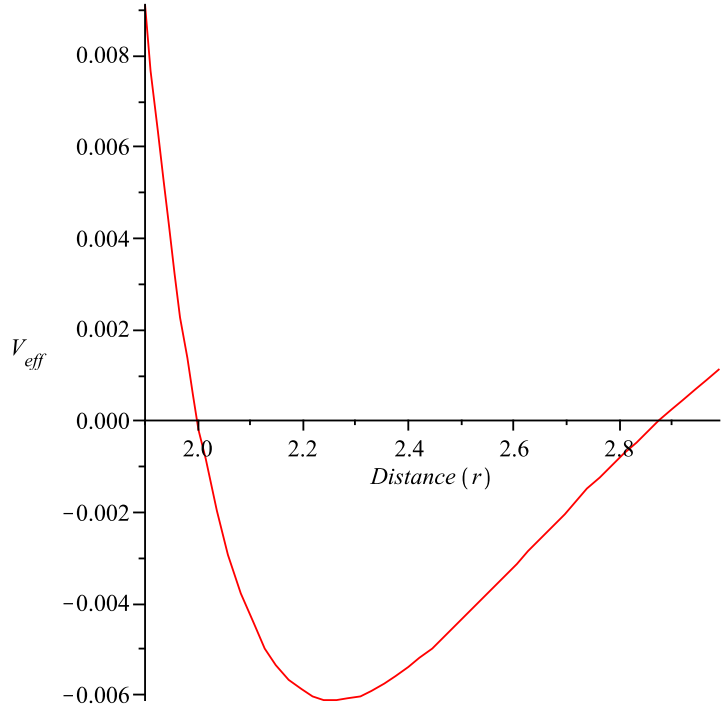}
    \caption{The effective potential for photons for M=2,G=1,$\beta=1$,$Q=\sqrt{2},l^2=80$,J=1. }
    \label{fig:Charged Brane black hole}
\end{figure}
Furthermore, there will be three sign changes in the $V_{eff}$.
Hence, there will be at most three positive roots for $V_{eff}$.
If we put E=0, then according to Descarte's rule of signs, the
effective potential has at most two positive roots.

\pagebreak

\begin{center}
\title{\bf 4.2  \textsc{For  Massive Particle ( L=1 )} }
\end{center}
The corresponding potential is given by,
\begin{equation}
 V_{eff} = - \frac{E^2}{2} + \frac{1}{2} f(r) \left( 1 +
 \frac{J^2}{r^2}\right)
\end{equation}
where
$f(r)=\left(1-\frac{2GM}{r}+\frac{Q^2+\beta}{r^2}+\frac{l^2Q^4}{20r^6}\right).$\\

It is to be mentioned that the roots of $f(r)$ are the horizons.
First, consider for radial geodesics with J = 0. As $f(r) > 0$ in
the region $ 0\leq r < r_-$, $V_{eff}$ will vanish for some
finite value of r in that region. Therefore, a time-like geodesic
will not reach the singularity. The massive particle will avoid
the singularity and would emerge in other regions. The space-time
is geodesically complete. We can analyze the various cases of
motion as follows:
If we take E=0, then $V_{eff}$ becomes
\begin{equation}
 V_{eff} =  \frac{1}{2}\left(1-\frac{2GM}{r}+\frac{Q^2+\beta}{r^2}+\frac{l^2Q^4}{20r^6}\right)
\end{equation}
\begin{figure}[htbp]
    \centering
       \includegraphics[scale=.5]{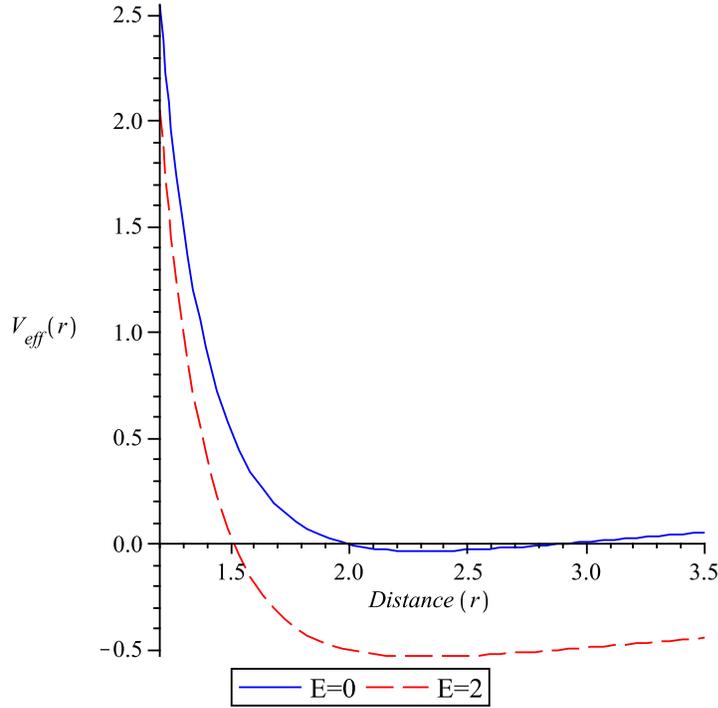}
    \caption{The effective potential (for radial geodesic) for massive particles for M=2,G=1,$\beta=1$,$Q=\sqrt{2},l^2=80$. }
    \label{fig:Charged Brane black hole}
\end{figure}

The zeros of the $V_{eff}$ coincide with the horizons. An example
of such a case is shown in Fig.7 .

From the shape of the potential, it is clear that the particle can
move only inside the black hole. Secondly, one can investigate the
behaviour of $V_{eff}$ for $ E \neq 0 $. The corresponding
$V_{eff}$ is given by eqn.(16). In this case, for $r \rightarrow
0$ the effective potential becomes
\[
V_{eff} \rightarrow \frac{l^2 Q^4}{20 r^6} +
\frac{Q^2+\beta}{r^2}- \frac{2GM}{r}
\]

For large r, $V_{eff} \rightarrow \frac{1-E^2}{2}$. For a black
hole with two horizons, in the two ranges, $0\leq r < r_- $ and
$r_+ < r $, the function $f(r) > 0$. Hence it is possible for
$V_{eff}$ to have roots in those two regions. Examples for two
roots are given in Fig.7.\\
Now we will consider the particles with angular momentum ( $ J
\neq 0 $). For E=0, the roots of the potential coincides with the
two horizons and the shape of the $V_{eff}$ is given in Fig. 8.
Hence the massive particle with "zero energy" would not escape
the black hole and would describe bounded orbits. The particle
could have circular stable orbits since the potential has a
minimum.\\

\begin{figure}[htbp]
    \centering
       \includegraphics[scale=.5]{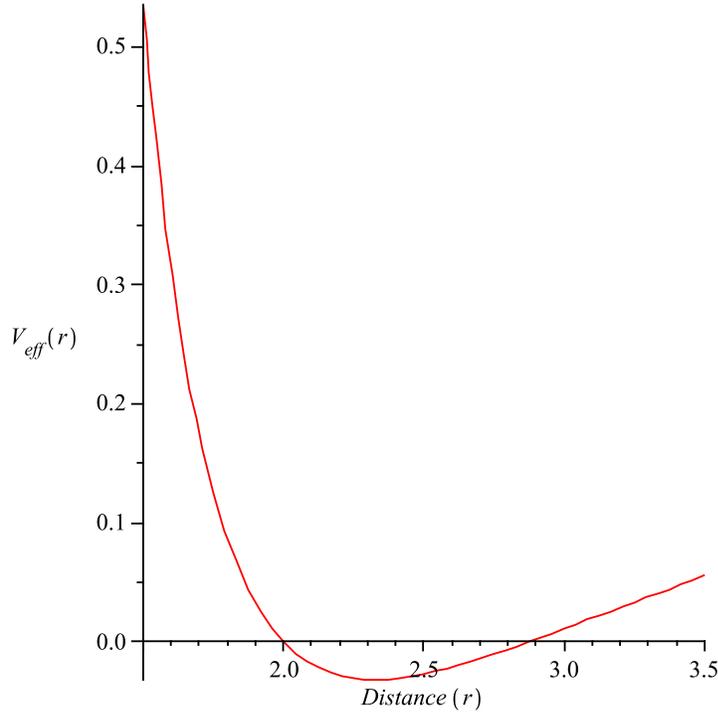}
    \caption{The effective potential for massive particles for M=2,G=1,$\beta=1$,$Q=\sqrt{2},l^2=80$,J=1. }
    \label{fig:Charged Brane black hole}
\end{figure}

Now for $ E \neq 0$, $V_{eff} \rightarrow  \frac{1-E^2}{2}$ for
large r.

 For $ r \rightarrow 0$,
\[
V_{eff} \rightarrow \frac{J^2}{2r^2}\left(1-\frac{2GM}{r}+
\frac{Q^2+\beta}{r^2}+\frac{l^2 Q^4}{20 r^6}\right).
\]
Similar to the arguments given for radial geodesics, it is
possible for $V_{eff}$ to have finite roots in the regions $0\leq
r < r_- $ and $r_+ < r $. $V_{eff}$ will have two or three roots
due to it's behaviour around $r = 0$. In both the cases a massive
particle would describe bounded orbits.\\

An example is shown in Fig.8 for two root cases. The two horizons
lie inside the region of the two roots of the potential. Hence
the particle will describe elliptic orbits. There is a minimum
for the potential as visible from Fig.8. Therefore it is possible
for a particle to have a stable circular orbit inside the black
hole.
\begin{center}
\title{\Large 5.  \textsc{Motion of test particle} }
\end{center}

Let us consider a test particle having mass,$m_0$ and charge, $ e
$ moving in the gravitational field of the Charged Brane-world
black hole described by the metric ansatz(1). So the
Hamilton-Jacobi [ H-J ] equation for the test particle is [4]
\[
g^{ik} \left( \frac{\partial S}{\partial x^i} + e A_i \right)
\left( \frac{\partial S}{\partial x^k} + e A_k \right) +m_0^2 =0
\]
where $g_{ik}$, \(A_i \)(gauge potential) are the classical
background fields (1)  and S is the standard Hamilton's
characteristic function.

For the metric (1) the explicit form of H-J equation  is  [5]
\begin{equation}
-\frac{1}{f} \left( \frac{\partial S}{\partial t} + \frac{e Q}{r}
\right)^2 + f \left( \frac{\partial S}{\partial r} \right)^2 +
\frac{1}{r^2}\left( \frac{\partial S}{\partial \theta}\right)^2 +
\frac{1}{r^2 \sin^2 \theta} \left( \frac{\partial S}{\partial
\phi } \right)^2 + m_0^2 =0
  \end{equation}

In order to solve this partial differential equation, let us
choose the $H-J$ function $ S $ as [5]
\[ S(t,r,\theta,\phi) = -E.t + S_1(r) + S_2(\theta) + {J. \phi} \]

 where $E$ is identified as the energy of the particle and $J$
 is the angular momentum of the particle.

 The radial velocity of the particle is ( for detailed
calculations, see $ref.[6]$ )
\begin{equation}
 \frac{ dr}{ dt} = f^2\left(E - \frac{eQ}{r^2}\right)^{-1}
 \sqrt{ \frac{1}{f^2} \left(E-\frac{eQ}{r} \right)^2 -
\frac{m_0^2}{f} - \frac{p^2}{fr^2} }
\end{equation}
where $p$ is the separation constant and termed as momentum of the
particle.

The turning points of the trajectory are given by
$\left(\frac{dr}{dt}\right) = 0 $ and we get

\[\left(E-\frac{eQ}{r} \right)^2 - m_0^2f - \frac{p^2}{r^2}f =0 \]
Solving
\[E= \frac{eQ}{r} + \sqrt{f} \left( m_0^2 + \frac{p^2}{r^2} \right)^{1/2} \]
The   potential curve is given by
\[ V(r) \equiv \frac{E}{m_0} = \frac{eQ}{m_0 r} + \sqrt{f} \left( 1+\frac{p^2}{m_0^2r^2} \right)^{1/2} \]
i.e. \[ V(r) = \frac{eQ}{m_0 r} +  \left( 1+\frac{p^2}{m_0^2r^2}
\right)^{1/2}\sqrt{1- \frac{2GM}{r}+\frac{Q^2 + \beta}{r^2} +
\frac{l^2 Q^4}{20 r^6}} \]

\pagebreak

In a stationary system, $ E $ i.e. $ V(r)$ must have an extremal
value. Hence the value of $r$ for which energy attains the
extremal value is given by the equation
\begin{equation}
         \frac{dV}{dr} =   0
          \end{equation}

which gives
\[
\frac{dV}{dr} = - \frac{eQ}{m_0 r^2} + \frac{1}{2\sqrt{f}}\left(1
+ \frac{p^2}{m_0^2r^2}\right)^{1/2}f^\prime (r) -  \sqrt{f}
\left(1+\frac{p^2}{m_0^2r^2}\right)^{-1/2} \frac{p^2}{m_0^2r^3} =0
\]

 We obtain
\[
\frac{eQ}{m_0r^2}\sqrt{f}\left( 1+ \frac{p^2}{m_0^2r^2}
\right)^{1/2} = \frac{1}{2} \left(1 + \frac{p^2}{m_0^2r^2} \right)
f^\prime (r) - f\frac{p^2}{m_0^2r^3}\]

Putting the expression of $f$ we obtain
\begin{equation}
\begin{split}
\frac{eQ}{m_0}\left( 1-\frac{2GM}{r}+\right. & \left. \frac{Q^2+\beta}{r^2} + \frac{l^2 Q^4}{20 r^6}\right)^{1/2} \left( 1 + \frac{p^2}{m_0^2r^2} \right)^{1/2} =  \\
 &\left(GM - \frac{Q^2 + \beta}{r} - \frac{3}{20}\frac{l^2 Q^4}{r^5} \right) \left(1 + \frac{p^2}{m_0^2r^2} \right)
 - \frac{p^2}{m_0^2 r} \left( 1 - \frac{2GM}{r} + \frac{Q^2+\beta}{r^2} + \frac{l^2 Q^4}{20r^6} \right)
 \end{split}
\end{equation}

\begin{center}
\title{\bf 5.1  \textsc{Test particle in Static Equilibrium} }
\end{center}

In static equilibrium, momentum $p$ must be zero. So, the value of
$r$ for which potential will be an extremal is given by
\[
\frac{eQ}{m_0} \left( 1 - \frac{2GM}{r} + \frac{Q^2 + \beta}{r^2}
+ \frac{l^2 Q^4}{20r^6} \right)^{1/2} = \left(GM - \frac{Q^2
+\beta}{r}- \frac{3}{20}\frac{l^2 Q^4}{r^5}\right)
\]
From this, we get
\[
\begin{split}
\left(G^2M^2 - \frac{e^2Q^2}{m_0^2} \right) r^{10}& + 2GM\left[\frac{e^2Q^2}{m_0^2}-(Q^2+\beta)\right]r^9+ (Q^2+\beta)\left[(Q^2+\beta)-\frac{e^2Q^2}{m_0^2}\right]r^8 \\
&-\frac{3}{10} GM l^2 Q^4 r^5 +
l^2Q^4\left[\frac{6(Q^2+\beta)}{m_0^2}-\frac{e^2Q^2}{20m_0^2}\right]
r^4 + \frac{9}{400} l^2 Q^8 =0
\end{split}
\]
If $ \frac{e^2Q^2}{m_0^2}> G^2M^2 $,we see that last term of the
equation is negative. So this equation has at least one positive
real root. Again, if $\frac{e^2Q^2}{m_0^2} = G^2M^2$ and
$\frac{e^2Q^2}{m_0^2} < (Q^2+\beta) $, then the above equation
changed to nine degree equation with negative last term implies a
real positive root exists.Therefore, it is possible to have bound
orbit for the test particle i.e. the test particle can be trapped
by the charged Brane-world black hole. In other words, the
charged Brane-world black hole exerts an attractive gravitational
force towards matter.

\begin{center}
\title{\bf 5.2  \textsc{Test particle in Non-Static Equilibrium } }
\end{center}

 \textit{\textbf{Case I : Uncharged test particle $(e=0)$}}

 Now the expression (21) simplifies to
$ $

 \[
\left(\frac{GM}{r^2} -\frac{Q^2+\beta}{r^3} -
\frac{3}{20}\frac{l^2Q^4}{r^7}\right)\left(1+\frac{p^2}{m_0^2r^2}\right)=\frac{p^2}{m_0^2r^3}
\left( 1 -\frac{2GM}{r} + \frac{Q^2+\beta}{r^2} + \frac{l^2
Q^4}{20 r^6}\right)
\]

$ $

 Thus we get the following algebraic equation as

\begin{equation}
 5 m_0^2 G Mr^7 - 5 m_0^2\left[ (Q^2+\beta) + \frac{p^2}{m_0^2} \right] r^6 +
 15p^2 GM r^5 - 10 p^2 (Q^2 +\beta)r^4
  - 15 l^2 Q^2 m_0^2 r^2 -p^2 l^2 Q^4 =0
\end{equation}

Obviously, this equation has at least one positive real root
since the last term of the above expression is negative. So it is
possible to have a bound orbit for the test particle. \\

 \textit{\textbf{Case II :  Test particle with charge $( e \neq 0 )$}}

From eqn.(21), we have the algebraic equation

\[
\begin{split}
\frac{eQ}{m_0r^2} &\left( 1 -\frac{2GM}{r} + \frac{Q^2+\beta}{r^2} + \frac{l^2 Q^4}{20 r^6} \right)^{1/2} \left( 1 + \frac{p^2}{m_0^2r^2} \right)^{1/2}\\
& = \left( \frac{GM}{r^2} - \frac{Q^2 +\beta}{r^3} - \frac{3}{20}
\frac{l^2 Q^4}{r^7} \right) \left(1 + \frac{p^2}{m_0^2r^2} \right)
- \frac{p^2}{m_0^2 r^3} \left( 1 - \frac{2GM}{r} +
\frac{Q^2+\beta}{r^2} + \frac{l^2 Q^4}{20 r^6} \right)
\end{split}
\]

$ $

 After simplifying the earlier eqn.  we get
\[
\begin{split}
 \left(G^2M^2 - \frac{e^2Q^2}{m_0^2} \right) r^{14}& + 2GM\left[\frac{e^2Q^2}{m_0^2}-(Q^2+\beta+\frac{p^2}{m_0^2})\right]r^{13}\\
 &+ \left[(Q^2+\beta+\frac{p^2}{m_0^2})^2
 +\frac{6p^2G^2M^2}{m_0^2}-\frac{e^2Q^2(Q^2+\beta)}{m_0^2}-\frac{e^2Q^2p^2}{m_0^4}\right]r^{12}\\
 &+ \left[\frac{2GMe^2p^2Q^2}{m_0^4}-\frac{6p^2GM(Q^2+\beta+\frac{p^2}{m_0^2})}{m_0^2}
 -\frac{4p^2GM(Q^2+\beta)}{m_0^2} \right]r^{11}\\
 &+\left[\frac{9G^2M^2p^4}{m_0^4}+\frac{4p^2(Q^2+\beta)(Q^2+\beta+\frac{p^2}{m_0^2})}{m_0^2}
 -\frac{e^2p^2Q^2(Q^2+\beta)}{m_0^4} \right]r^{10}\\
&-\left[\frac{3GMl^2Q^4}{10} +\frac{12p^4GMQ^2(Q^2+\beta)}{m_0^4} \right]r^{9}\\
&+\left[\frac{4p^4(Q^2+\beta)^2}{m_0^4}+\frac{6l^2Q^4(Q^2+\beta+\frac{p^2}{m_0^2})}{m_0^2}
 -\frac{e^2l^2Q^6}{20m_0^2} \right]r^{8}
  -\frac{13  p^2GM l^2 Q^4}{20 m_0^2} r^7 \\
  &+\left[-\frac{e^2l^2p^2Q^6}{20m_0^4}+\frac{2p^2l^2Q^4(Q^2+\beta+\frac{p^2}{m_0^2})}{5m_0^2}
 +\frac{12p^2l^2Q^4(Q^2+\beta)}{20m_0^2}
 \right]r^{6}\\
 &-\frac{6GMl^2p^4Q^4}{5m_0^4}r^5+\left[\frac{9l^2Q^8}{400}+\frac{4p^4l^2Q^4(Q^2+\beta)}{5m_0^4}\right]r^4
+\frac{6p^2l^2Q^8}{100m_0^2}r^2 + \frac{p^4l^4Q^8}{25m_0^4}  =0
\end{split}
\]

If $ \frac{e^2Q^2}{m_0^2}> G^2M^2 $,we see that last term of the
equation is negative. So this equation has at least one positive
real root. Again,if $\frac{e^2Q^2}{m_0^2} = G^2M^2$ and
$\frac{e^2Q^2}{m_0^2} < (Q^2+\beta+\frac{p^2}{m_0^2}) $, then the
above equation changed to thirteen degree equation with negative
last term implies a real positive root exists.Therefore, it is
possible to have bound orbit for the test particle i.e. the test
particle can be trapped by the charged Brane-world black hole. In
other words, the charged Brane-world black hole exerts an
attractive gravitational force towards matter in this case also.
\begin{center}
\title{\Large 6.  \textsc{ Thermodynamics} }
\end{center}
For a static black hole, Hawking temperature $T_H$ is an
important thermodynamical quantity. For the Charged brane-world
black hole metric $T_H$ is given by
\[
T_H =
\frac{1}{\sqrt{-g_{tt}g_{rr}}}\frac{d}{dr}(-g_{tt})\mid_{r=r_{h}}
 \]
Now $g_{tt}=f(r)= 0 $ yields ( See details in Annexure )
\begin{equation}
r^3 + a_1 r^2 + a_2 r + a_3 = 0
\end{equation}
 where $ a_1 = - \left[ GM +
\sqrt{G^2M^2-Q^2-\beta}\right]$ ;
 $ a_2 =\frac{lQ^2}{\sqrt{20}\sqrt{G^2M^2-Q^2-\beta}}$ ;
 $ a_3 = -\frac{lQ^2}{\sqrt{20}} $ .\\

A straightforward analysis shows that there are three possible
cases for Eqn.(23). The first case corresponds to
\[
12 a_2^3 + 81 a_3^2 + 12 a_3 a_1^3 < 3 a_2^2 a_1^2 + 54 a_1 a_2
a_3
\]
for which Eqn.(23) has no real root. So the singularity is
naked.\\

The second case corresponds to
\[
12 a_2^3 + 81 a_3^2 + 12 a_3 a_1^3 = 3 a_2^2 a_1^2 + 54 a_1 a_2
a_3
\]
and has one real positive root, which corresponds to an extremal
black hole.\\

Finally, if
\[
12 a_2^3 + 81 a_3^2 + 12 a_3 a_1^3 > 3 a_2^2 a_1^2 + 54 a_1 a_2
a_3 , \]
 there are two real positive roots and the black hole has
both an outer and inner horizon.

 Obviously, the roots of Eqn.(23) are given by
\[ r = r_h = S + T -
\frac{a_1}{3} \]
 With $ S = \sqrt[3]{R + \sqrt{P^3 + R^2}} $, $ T
= \sqrt[3]{R - \sqrt{P^3 + R^2}} $ \\

where  $P = \frac{3 a_2 - a_1^2}{9}$; $ R = \frac{9 a_1 a_2 - 27
a_3 - 2a_1^3}{54} $.\\

Using Eqn.(1) the Hawking temperature becomes
\[
T_H = \frac{1}{2\pi}\left[ \frac{GM}{r_h^2}-\frac{Q^2 +
\beta}{r_h^3} - \frac{3 l^2 Q^4}{20 r_h^7}\right]
\]
with $r_h$ is the location of the (outer) event horizon. If $GM=
\frac{Q^2 + \beta}{r_h} + \frac{3 l^2 Q^4}{20 r_h^5}$, then $T_H = 0 $. Thus in that case they are stable end points of Hawking evaporation.\\

Also, the entropy is given by $ S = \frac{(area)}{4} = \pi r_h^2 $
and the surface gravity is given by
\[
\chi = \frac{1}{2} \left[\frac{\partial f(r)}{\partial
r}\right]_{r=r_h} = \left[ \frac{GM}{r_h^2}-\frac{Q^2 +
\beta}{r_h^3} - \frac{3 l^2 Q^4}{20 r_h^7}\right] .
\]

\begin{center}
\title{\Large 7.  \textsc{ Solution of massless Scalar Wave Equation in Charged
brane-world black hole metric} }
\end{center}

Here, we shall analysis the scalar wave equation for charged
brane-world black hole geometry following Brill et. al [7]. The
wave equation for a massless particle is given by

\[
g^{-1/2} \frac{\partial }{\partial x^\mu} \left( g^{1/2}
g^{\mu\nu} \frac{\partial }{\partial x^\nu} \right) \chi  =0
\]
Here $g_{\mu\nu} $is given by Eq. (1). Putting all the values we
get

\[
-\frac{r^4 \sin\theta}{\Delta}\frac{\partial^2 \chi}{\partial t^2}
+ \sin \theta \frac{\partial }{\partial r}\left(\Delta
\frac{\partial \chi}{\partial r}\right) + \frac{\partial
}{\partial \theta}\left( \sin\theta \frac{\partial }{\partial
\theta} \right)\chi + \frac{1}{\sin\theta}\frac{\partial^2
\chi}{\partial \phi^2}=0
\]
Where $\Delta = r^2- 2G Mr + (Q^2 +
\beta)+\frac{l^2Q^4}{20 r^4}$. \\

This equation can be solved by using separation of variable with
the ansatz
\[ \chi = e^{-i\omega t} e^{im\phi} R(r) \Theta(\theta) \]
Substituting  this in the wave equation we get
\[
\frac{r^4 \sin \theta}{\Delta}\omega^2 \chi + \frac{\sin
\theta}{R} \frac{\partial }{\partial r}\left(\Delta
\frac{\partial R}{\partial r}\right)\chi + \frac{1}{\Theta}
\frac{\partial }{\partial \theta}\left(\sin\theta\frac{\partial
\Theta }{\partial \theta} \right)\chi - \frac{m^2}{\sin\theta}
\chi =0
\]

 The radial equation reduces to

\[
\Delta \frac{\partial }{\partial r}\left( \Delta \frac{\partial
R}{\partial r} \right) + (r^4 \omega^2 - \Delta \lambda)R=0
\]

The angular part becomes
\[
\frac{1}{\sin\theta}\frac{\partial}{\partial \theta}
\left(\sin\theta\frac{\partial \Theta}{\partial \theta}\right) -
\frac{m^2}{\sin^2\theta} \Theta + \lambda \Theta =0
\]

 Substituting $x=\cos\theta$, the equation becomes

\[
(1-x^2)\frac{d^2 \Theta }{ dx^2} - 2x \frac{d\Theta}{dx} - \left(
\frac{m^2}{1-x^2} - \lambda \right) \Theta =0
\]

If we now write $\lambda = l(l+1)$ where $l$ is an integer  then
the equation
\[
(1-x^2)\frac{d^2 \Theta}{dx^2} - 2x \frac{d \Theta}{dx} + \left(
l(l+1) - \frac{m^2}{1-x^2}  \right) \Theta =0
\]
is Associated Legendre equation and the solution is given by the
associated Legender polynomial $ P^m_l(x)$ and is expressed as
\[
\Theta^m_l(\cos\theta) = P^m_l(x) = \frac{(-1)^m}{2^l l!}\left(
1-x^2 \right)^{m/2} \frac{d^{l+m} }{dx^{l+m}} \left( x^2 -1
\right)^{l}
\]

\begin{center}
\title{\bf 7.1  \textsc{The Radial Equation:Absence of Superradiance} }
\end{center}
No energy extraction is possible for Schwarzschild black hole.
Whereas Kerr-Newman black hole allows energy extraction. An
explicit process (Penrose process) by which this can be achieved
was first outlined by Roger Penrose in 1969. Superradiance is
nothing but the wave analogue of the Penrose process on Black
Hole. If a bosonic or fermionic wave is incident upon a black
hole, normally the reflected wave carries less energy than the
incident wave. But under certain condition the transmitted wave,
absorbed by the black hole carries negative energy into the black
hole making the reflection coefficient for the wave greater than
unity. That implies that the reflected wave will carry more energy
than the incident wave. This phenomenon is called superradiance by
Misner [15] and also analyzed by Zel'dovich and Starobinsky [10,
11]. Through this process energy can be extracted from a black
hole in expense of its angular momentum. The condition is given by
[12]
\[
     0 < \omega < m \Omega_H,
\]
where $\Omega_H$ is the angular velocity of the horizon [9]. By
considering Kerr geometry,  Chandrasekhar [8] has analyzed this
phenomenon and has  shown that this phenomenon occurs only for
incident waves of integral spins, i.e., for scalar,
electromagnetic waves and gravitational cases. Also, he has shown
its absence for the fermionic waves, i.e. Dirac wave or neutrino
waves. Basak and Majumdar discussed this phenomenon for acoustic
analogue of Kerr Black Hole [13, 14]. It is argued that
superradiance phenomenon is possible if the black hole rotates or
is charged [16].  Now we check whether the superradiance
phenomenon will happen for charged brane-world black hole.

The radial equation is given by
\[
\Delta \frac{d}{dr}\left( \Delta \frac{ dR}{dr} \right) + \left(
\omega^2 r^4 - l(l+1)\Delta\right)R=0
\]

Let us introduce the familiar  $ r^\ast $ coordinate (the tortoise
coordinate) defined by
\[
\frac{d r^\ast}{ d r} = \frac{r^2}{\Delta},
\]
thus giving
\[
\Delta \frac{ d }{ d r} = r^2 \frac{d}{ d r^\ast}.
\]

 Note that though the variable $r^\ast$  is defined in the same manner as
in Schwarzschild or in Kerr metric, in this case the variable is
non-integrable. Still the basic purpose is satisfied, the
coordinate spans over the real line and pushes the horizon to
minus infinity.

The introduction of another function $u(r) = rR$ reduces the
radial equation as
\[
\frac{d^2 u}{ {d }r^{\ast 2}} + \left[ \frac{2 \Delta^2}{r^6} -
\frac{\Delta}{r^5} \frac{ d\Delta }{dr} -\frac{ \lambda
\Delta}{r^4} + \omega^2\right] u = 0
\]
Putting the value of $d\Delta/dr$ we get
\[
\frac{d^2 u}{ {d }r^{\ast 2}} + \left[ \frac{l^2Q^4 \Delta}{5
r^{10}}+ \frac{2 \Delta^2}{r^6} + \frac{2 M \Delta}{r^5}-
 \left(2+l(l+1)\right)\frac{\Delta}{r^4}+ \omega^2\right] u = 0
\]

Thus a potential barrier remains where
\[
V(r) = -\left[ \frac{l^2Q^4 \Delta}{5 r^{10}}+ \frac{2
\Delta^2}{r^6} + \frac{2 M \Delta}{r^5}-
 \left(2+l(l+1)\right)\frac{\Delta}{r^4}+ \omega^2\right]
\]

At horizon ($\Delta \rightarrow 0, r^\ast \rightarrow - \infty$),
the radial equation becomes
\[
\frac{d^2 u_H}{ dr{\ast^2}} + \omega^2 u_H =0
\]
with $V(r) = - \omega^2$.

Now asymptotically, $r \rightarrow \infty $ implying $ r^\ast
\rightarrow \infty$.
 The equation has the same form as in
the previous case
\[
\frac{d^2 u_\infty}{dr^{\ast2}} + \omega^2 u_\infty =0
\]

Thus $u_H = u_\infty$, where $u_H$ is the radial solution at
horizon and $u_\infty$ is the solution at $\infty$. This equality
shows that for a charged Brane-world black hole metric there is
\textit{no phenomenon of superradiance for an incident massless
scalar field}.

\pagebreak

\begin{center}
\title{\Large 8.  \textsc{ Concluding Remarks} }
\end{center}
In the present investigation, we have analyzed the behavior of the
time-like and null geodesics of the charged brane-world black
holes. Two types of charge can arise on the brain, one from the
bulk Weyl tensor and another from a Maxwell field trapped on the
brane. Figures (1) and (3) indicate that the nature of ordinary
time w.r.t. radial distance for the massless and massive particle
in the gravitational field of charged brane-world black hole have
the same nature. Here, one can see that ordinary time increases
with increase of radial distance.  Figures (2) and (4) shows the
similar kind of nature for proper time-distance graph. For radial
geodesics, the effective potential for massless particle is
independent on the charge and mass of the black hole where as from
the shape of potential, it is clear that the time-like particle
can move only inside the black hole. For circular geodesics, the
roots of the effective potential coincide with the horizon and
also as the potential has a minima between the horizons, the
photon-like as well as time-like particles would be
bounded in a stable circular orbit.\\
In this paper, we also investigate the motion of test particles in
the gravitational field of charged brane-world black holes  using
Hamilton-Jacobi ( H-J ) formalism. The test particle is considered
to be both static and non-static as well as charged or
uncharged. \\
In static case, we have seen that the test particle can be trapped
by the charged brane-world black hole provided that either $
\frac{e^2Q^2}{m_0^2}> G^2M^2 $ or  ( $\frac{e^2Q^2}{m_0^2} =
G^2M^2$ and $\frac{e^2Q^2}{m_0^2} < (Q^2+\beta) $ ). For
non-static equilibrium, uncharged  test particle always be trapped
where as charged test particle can be trapped provided that either
$ \frac{e^2Q^2}{m_0^2}> G^2M^2 $ or ( $\frac{e^2Q^2}{m_0^2} =
G^2M^2$ and $\frac{e^2Q^2}{m_0^2} <
(Q^2+\beta + \frac{p^2}{m_0^2}) $ ).\\
 It is known that
Superradiance phenomenon could be  seen in charged or rotating
black holes [16 ]. In this study, we have shown that Superradiance
phenomenon is absent in charged brane-world black hole.\\
 We have  also studied the thermodynamics of the
charged brane-world black hole. One can see that the charged
brane-world black hole exhibits a non zero entropy at zero
temperature under a certain condition, say, $GM= \frac{Q^2 +
\beta}{r_h} + \frac{3 l^2 Q^4}{20 r_h^5}$. Also at this particular
situation, the surface gravity would vanish. It is observed that
mass of the charged brane-world black hole plays a crucial role to
increase the horizon in other words, to increase the entropy.
Finally, one can note that for $M=0$, the solution (1) describes a
naked singularity.

\begin{center}
\title{\Large  \textsc{ Acknowledgments} }
\end{center}

          MK has been partially supported by UGC, Government of India under MRP scheme.
          FR is thankful to DST , Government of India for providing financial support.
          BR likes to thank The Inter-University Centre for Astronomy and Astrophysics (IUCAA), Pune, India for hospitality during a visit
          under the Associateship programme.\\

\pagebreak

\begin{center}
\title{\Large   \textsc{ Appendix} }
\end{center}

 At the horizon ($r=r_h$), $f(r) =0$ implies
\[
1- \frac{2GM}{r}+\frac{Q^2 + \beta}{r^2} + \frac{l^2 Q^4}{20r^6}=0
\]
 i.e.
\begin{equation}
r^6 - 2GMr^5 + (Q^2 + \beta)r^4 + \frac{l^2Q^4}{20} = 0
\end{equation}

One can write the above equation as
\begin{equation}
 (r^3  - GMr^2 + A )^2 - ( Br^2 + Cr +D )^2  = 0
\end{equation}
After simplification we get
\begin{equation}
r^6 - 2GMr^5 + (G^2M^2 - B^2)r^4 + (2A-2BC)r^3 -(2GMA+2BD+C^2)r^2
- 2CD r + (A^2 - D^2) = 0
\end{equation}
Comparing eqn.(24) and eqn.(26) we get
\begin{equation}
G^2M^2 - B^2 = Q^2 + \beta
\end{equation}
\begin{equation}
 2A-2BC  = 0
\end{equation}
\begin{equation}
 2GMA+2BD+C^2  = 0
\end{equation}
\begin{equation}
- 2CD   = 0
\end{equation}
\begin{equation}
A^2 - D^2 = \frac{l^2Q^4}{20}
\end{equation}
Eqn.(30) implies either $ C =0 $ or $ D =0$. But eqn.(28) implies
if $ C=0 $, then $ A=0 $. Therefore, we take $ C \neq 0 $ and $
D=0$.\\

Now, eqn.(27),(31) implies $B = \sqrt{G^2M^2-Q^2-\beta}$ and $A =
-\frac{lQ^2}{\sqrt{20}}$ \\
Again, Eqn.(28) implies
\[
C=A/B=-\frac{lQ^2}{\sqrt{20(G^2M^2-Q^2-\beta)}}
\]
Consistency Condition :  Putting all values in eqn.(29), we get
\[
2GMBC + C^2 =0 \Rightarrow 2GMB^2 + A =0
\]
i.e.
\[
2GM(G^2M^2-Q^2-\beta)=\frac{lQ^2}{\sqrt{20}}
\]
Now, from eqn.(25) we get
\[
r^3  - GMr^2 + A = \pm( Br^2 + Cr )
\]
Taking only $+ ve$ sign, we get
\begin{equation}
r^3  - (GM +B)r^2 - Cr + A = 0
\end{equation}
Here, we note that $ A < 0$ and $ C < 0$. Hence, eqn.(32) has
three changes of sign. From the Descarte's rule of sign, the above
equation has at most three $+ ve$ roots. Putting all the values of
B,C,A, we get
\[
r^3 + a_1 r^2 + a_2 r + a_3 = 0
\]
 where $ a_1 = - \left[ GM +
\sqrt{G^2M^2-Q^2-\beta}\right]$ ;
 $ a_2 =\frac{lQ^2}{\sqrt{20}\sqrt{G^2M^2-Q^2-\beta}}$ ;
 $ a_3 = -\frac{lQ^2}{\sqrt{20}} $ .\\

Obviously, the roots of the above  equation are given by
\[ r = r_h = S + T -
\frac{a_1}{3} \]
 With $ S = \sqrt[3]{R + \sqrt{P^3 + R^2}} $, $ T
= \sqrt[3]{R - \sqrt{P^3 + R^2}} $ \\
where  $P = \frac{3 a_2 - a_1^2}{9}$; $ R = \frac{9 a_1 a_2 - 27
a_3 - 2a_1^3}{54} $.\\

For Graphical representation :\\
Values are taken following the Consistency Condition as $ G =
\beta =1, M=2, Q=\sqrt{2}, l^2=80$. Hence, the equation becomes
\[
f(r)\equiv r^6 - 4r^5 + 3r^4 + 16 \equiv 0
\]
The roots are shown in Fig.9.
\begin{figure}[htbp]
    \centering
       \includegraphics[scale=.3]{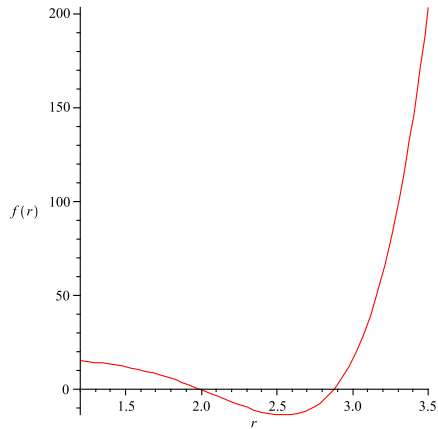}
    \caption{The roots of $f(r)$ for M=2,G=1,$\beta=1$,$Q=\sqrt{2},l^2=80$. }
    \label{fig:Charged Brane black hole}
\end{figure}


\begin{thebibliography}{99}
\bibitem{kg01} N. Dadhich, R. Maartens, P. Papadopolous  and  V. Rezania,   Phys. Lett. B487, 1(2000).
\bibitem{kg02} A. Chamblin, H.S. Reall, H.A. Shinkai and T. Shiromizu,  Phys.Rev.D63, 064015(2001).
\bibitem{kg03} T. Shiromizu, K. Maeda and M. Sasaki,  Phys.Rev.D62, 024012(2000).
\bibitem{kg04} L. D. Landau, The Classical Theory of Fields, (Pergamon, 1973).
\bibitem{kg05} F. Rahaman, Int. J. Mod. Phys. D, 9(5), 627-632
(2000) ; F. Rahaman et al, Czech.J.Phys.53,115(2003).

\bibitem{kg06} S. Chakraborty, Gen. Rel. Grav. 28, 1115(1996); \\
               S. Chakraborty, F. Rahaman, Pramana 51, 689(1998).
\bibitem{kg07} D. R. Brill, P. L. Chrzanowski, C. M. Pereira, E. D. Fackerell, and J. R. Ipser, Phys. Rev. D, 5, 1913 (1972).
\bibitem{kg08} S. Chandrasekhar,  The Mathematical Theory of Black Holes.
\bibitem{kg09} Bryce. s. DeWitt, Quantum Field Theory in Curved Spacetime, Phys. Rep., 19(6),p.295-357, Aug 1975.
\bibitem{kg10} Ya. B. Zel'dovich, Soviet Phys.– JEPT, 35, 1085–1087, (1972).
\bibitem{kg11} A. A. Starobinskii, Soviet Phys.– JEPT, 37, 28–32, (1973).
\bibitem{kg12} R. M. Wald, General Relativity, (Overseas India, New Delhi, 2006).
\bibitem{kg13} S. Basak, P. Majumdar, Class.Quant.Grav. 20, 2929-2936 (2003).
\bibitem{kg14} S. Basak, P. Majumdar, Class.Quant.Grav. 20, 3907-3914 (2003).
\bibitem{kg14} C. Misner, Phys.Rev.Lett.28, 994 (1972)
\bibitem{kg14} K Shiraishi, Mod.Phys.Lett.A, 37, 3449 (1992); M H
Ali, Gen.Rel.Grav., 37, 977 (2007).
\end{thebibliography}
\end{document}